\documentclass[12pt]{iopart}
\usepackage{tensor}
\usepackage{iopams}
\usepackage{float}
\usepackage{tikz}
\usepackage{graphicx}
\usepackage{float} 
\usepackage{subcaption}
\usetikzlibrary{backgrounds}
\usepackage{subcaption}
\newcommand{\eqref}[1]{(\ref{#1})}

\begin{document}
\title[Revisiting the gravitational ``arrow of time'']{Revisiting the gravitational ``arrow of time''}
\author{R A Sussman$^1$, S N\'ajera$^2$, F A Piza\~na$^1$, J C Hidalgo$^2$}

\address{$^1$ Instituto de Ciencias Nucleares, Universidad Nacional Aut\'onoma de M\'exico
(ICN--UNAM),\\ A. P. 70 –- 543, 04510 M\'exico D. F., M\'exico.}
\ead{sussman@nucleares.unam.mx}

\address{$^2$ Instituto de Ciencias F\'isicas, Universidad Nacional Aut\'onoma de M\'exico, 62210, Cuernavaca, Morelos.}
\ead{sebastian.najera@icf.unam.mx}

\begin{abstract}
We address a long-standing misperception on the gravitational ``arrow of time'', a proposal by Penrose (also known as the ``Weyl-curvature hypothesis") that associates structure formation along timelike directions in which Weyl-curvature scalars become dominant over Ricci scalars. A counterexample of this hypothesis was found by Bonnor on a class of exact solutions describing heat conducting spheres collapsing in a Vaidya background. We show that this result does not hold in the same class of solutions considered as physically viable near FLRW cosmological models, with the heat conduction vector interpreted as a peculiar velocity field. We also discuss the similarities and differences between the gravitational ``arrow of time'' and the gravitational entropy formalism of Clifton, Ellis and Tavakol. 
\end{abstract}

\section{Introduction}\label{intro}

The nearly homogeneous early Universe evolved with entropy production driven by thermal processes close to thermal equilibrium. Penrose \cite{penrose1979singularities} argued that a mechanism was required to guide, in a timelike direction, self-gravitating systems that undergo structure formation from an early Ricci-curvature dominant universe to a Weyl-curvature scenario. This led to the concept of a gravitational ``arrow of time'', later formalized in terms of the proper time evolution of a ratio between the Weyl and Ricci tensors \cite{wainwright1984,goode1985}. This notion has motivated a published  literature with discussion, criticism and various appraisals \cite{pelavas2006gravitational,gron2002weyl,Rudjord:2006v,gron2012entropy,gregoris2020thermodynamics,chakraborty2021investigation,gregoris2022understanding,chakraborty2024arrow, zhao2018black, guha2020gravitational}. In particular, \cite{chakraborty2024arrow} employed the Kretschmann scalar to define a ratio of Weyl to Riemann curvatures applicable to vacuum spacetimes. 

While this curvature ratio might provide an appealing  geometrical characterization of gravitational clustering, its proper theoretical interpretation has been subject of debate. Bonnor \cite{bonnor1985gravitational} compared Penrose's gravitational arrow with a "thermodynamic arrow of time" defined by the direction of increasing matter inhomogeneity. He showed that both arrows do not coincide in shear-free heat conducting fluids where Penrose's proposal decreases with proper time, see Sec.~\ref{sec:PenBon} for details. This was confirmed in later works \cite{ystein2008weyl, chakraborty2021investigation, chakraborty2024arrow}, apparently challenging Penrose's hypothesis. However, Bonnor's analysis was restricted to collapsing configurations constrained by exterior matching to Vaidya spacetimes. 
Such matchings have been extensively studied \cite{de1985collapse,santos1985non, kolassis1988energy, bonnor1989radiating,grammenos1995radiating} modeling spheres as shear-free conducting fluids by relying on highly simplified metric ansatzes. These constructions aimed to interpret the heat flux in terms of constitutive equations of non-equilibrium thermodynamics \cite{maharaj2012radiating, schafer2000gravitational, tewari2015dissipative}. However, when the same class of shear-free solutions is reinterpreted as cosmological models without such boundary conditions, Penrose's prediction is recovered. 

In this work, we revisit this apparent contradiction by showing that when the shear-free solutions are considered as expanding cosmological models without an exterior matching, the Weyl to Ricci ratio evolves in the expected direction, consistent with the gravitational arrow of time. We further analyze this behaviour in terms of physically viable models by the reinterpretation of heat flux as peculiar velocities. The section by section content of the paper is as follows: Section \ref{sec:PenBon} discusses both Penrose's and Bonnor's gravitational entropy proposal, in Section \ref{sec:gravarrow} we provide a plausible cosmological model as a counterexample to Bonnor's argument that Penrose's gravitational arrow of time must evolve in an opposite direction. Finally in Section \ref{sec:closing} we present our closing remarks. 

\section{Comparing Penrose and Bonnor's gravitational entropy proposal}\label{sec:PenBon}

Penrose's qualitative notion of an ``arrow of time'' was formalized as the proper time evolution of the following dimensionless ratio \cite{wainwright1984,goode1985} 
 \begin{equation} \hbox{as}\,\,\tau \,\,\hbox{increases}\quad {\cal P} = \frac{C_{abcd}C^{abcd}}{R_{ab} R^{ab}}\quad\hbox{increases:} \quad  \dot {\cal P}\geq 0,\label{ratioWR}\end{equation}
where the dot is $u^a \nabla_a$. This notion has triggered a literature of discussion, criticism and various appraisals \cite{pelavas2006gravitational,gron2002weyl,Rudjord:2006v,gron2012entropy,gregoris2020thermodynamics,chakraborty2021investigation,gregoris2022understanding,chakraborty2024arrow}. In particular, \cite{chakraborty2024arrow} also used the Kretschmann scalar as a ratio of Weyl to Riemann curvatures applicable to vacuum spacetimes.

 Bonnor \cite{bonnor1985gravitational} defined a gravitational entropy current as ${\cal P}^a = {\cal P}\,u^a$ and proved that in known dust cosmological models (LTB and Szekeres \cite{Bonnor1986}) the ``gravitational arrow'' condition \eqref{ratioWR} points in the same direction as (what he defined as) the ``arrow of time'' along which structure formation develops. Yet, when looking at the collapse of shear-free heat conducting spheres matched to a Vaidya exterior Bonnor found \cite{BONNOR1987305} the two arrows pointing in opposite directions: as the spheres collapse, the outgoing null radiation increases (arrow of time), while the ratio \eqref{ratioWR} of the gravitational arrow decreases. This result was validated recently by Chakraborty et al \cite{chakraborty2024arrow}.  

However, as we show in this paper, Bonnor's result is not generic to the class of spherically symmetric, shear-free, heat conducting solutions, but a direct consequence of the matching conditions in their collapse in a Vaidya background. Such collapse involves outgoing radiation into the Vaidya background, and is the only physically viable situation (an expanding sphere absorbing radiation from the Vaidya exterior would be unphysical). The monotonic decrease of the ratio \eqref{ratioWR} can be readily appreciated from the Vaidya exterior solution itself, whose metric and coherent radiation energy-momentum are
\begin{eqnarray}  ds^2&=& -\left(1-\frac{2M(u)}{r}\right)du^2 +2 du dr + r^2(d\theta^2+\sin^2\theta d\phi^2),\label{vaidya}\\
T^{ab} &=&\Phi(u)\, l^a l^b,\label{TabV}
\end{eqnarray}
where $l^a=\delta^a_v$ is a null vector, $\Phi$ is the radiation energy density and $M(u)$ is the varying mass, which is necessarily a decaying function $dM/du<0$ because energy conditions require $\Phi(u)\geq 0$. Einstein's equations reduce to
\begin{equation}R^{ab}=-\frac{2 M_{,u}}{r^2}\,l^a l^b =8\pi \Phi(u)\, l^a l^b \end{equation}
with $R=g_{ab}R^{ab}=0$ and $R_{ab}R^{ab}=0$ (and $R_{abcd}R^{abcd}=C_{abcd}C^{abcd}$), making it impossible to form the Weyl to Ricci ratio as in \eqref{ratioWR}. However, an analogous ratio can be formed with invariants of the Weyl and Ricci tensors \cite{RiemannInvars}: the unique conformal Weyl invariant $\Psi_2$ and the unique component of the trace-free Ricci spinor $\Phi_{22}$, which for the Petrov type D Vaidya solution \cite{coudray2021geometry} leads to a ratio of Weyl to Ricci curvatures: 
\begin{equation}\Psi_2 = -\frac{M}{r^3},\qquad \Phi_{22} =  -\frac{M_{,u}}{r^2}, \quad \Rightarrow\quad \frac{\Psi_2}{\Phi_{22}}=\frac{M}{r\,M_{,u}}.\label{vaydia2} \end{equation}
Considering the asymptotic structure of the Vaidya solution in the limit $u\to\infty$ \cite{VaidyaExtension}, we clearly have 
\begin{equation}\lim_{u\to\infty} \frac{\Psi_2}{\Phi_{22}}=\lim_{u\to\infty}  \frac{M}{r\,M_{,u}}=\infty,\end{equation}
since $r\to 2M(u)$ and $M_{,u}\to 0$ as $u\to \infty$. Therefore, in the Vaidya exterior Ricci curvature dominates Weyl-curvature as the radiation emitted by the null dust sources of the Vaidya solution reaches the future event horizon $r= 2M(u)$. 

The heat conducting fluid model used by Bonnor \cite{BONNOR1987305} and other authors \cite{de1985collapse, santos1985non, kolassis1988energy, bonnor1989radiating, grammenos1995radiating, chakraborty2024arrow} is described in detail in their articles. The metric, energy-momentum tensor and field equations are  
\begin{eqnarray} 
ds^2=-A_0^2(\chi) dt^2 + a^2(t) B_0^2(\chi) [d\chi^2+\chi^2(d\theta^2+\sin^2\theta d\phi^2)],\label{metric1}\\
T^{ab}=\rho u^a u^b + ph^{ab} + 2q^{(a}u^{b)},\qquad u^a = A_0^{_1}\delta^a_t,\quad q_a=Q\,\delta_a^\chi,\label{Tab}
\end{eqnarray}    
\begin{eqnarray}
\fl 8\pi Q = -\frac{2 a_{,t}}{a}\,\frac{A_0'}{A_0^2},\qquad 
\ \frac{8\pi}{3}\rho=\frac{8\pi \rho_0}{3a^2}+\frac{1}{A_0^2}\frac{ a_{,t}^2}{a^2},\qquad
8\pi p = \frac{8\pi p_0}{a^2}-\frac{1}{A_0^2}\left(\frac{a_{,t}^2}{a^2}+\frac{2a_{,tt}}{a}\right),\nonumber\\\label{Qrhop}
\end{eqnarray}
where $A_0=A_0(\chi)$ and $B_0=B_0(\chi)$ must comply with the condition of pressure isotropy $G^\chi_\chi-G^\theta_\theta=0$ (Equation (33) of \cite{chakraborty2024arrow}). Bonnor (and subsequent authors) assumed that $\rho_0=\rho_0(r)$ and $p_0=p_0(r)$ (given by equations (4.4) and (4.5) of \cite{bonnor1989radiating}) correspond to an asymptotically initial static fluid with $a_{,t}=0$ and $q^a= 0$ at $t\to-\infty$ \footnote{Practically all work on these models use this simplified metric, but the most general solution (without isometries) for an irrotational shear-free fluid whose source is \eqref{Tab} was derived in \cite{sussman1993new}}. 

For Petrov type D models described by \eqref{metric1}-\eqref{Tab} the Weyl tensor reduces to $C_{abcd}C^{abcd} = 8E_{ab}E^{ab}=8\Psi_2^2$, therefore, we can write \eqref{ratioWR} in the fluid region as the following dimensionless ratio
\begin{equation}{\cal P}=\frac{8E_{ab}E^{ab}}{R_{ab}R^{ab}}=\left(\frac{3H_0^2 c^4}{8\pi G}\right)^2\, \frac{\frac43 \Psi_2^2}{ \rho^2+3p^2-2q_aq^a }.\label{ratioWR2}\end{equation}
which must be evaluated in consistency with the field equations and the matching conditions with Vaidya that in the fluid region is a comoving boundary $\chi=\chi_\Sigma$.      

For the purpose of this paper we only need the following matching condition for \eqref{metric1}-\eqref{Tab} with a Vaidya exterior taking place along a fixed $\chi=\chi_\Sigma$ (see detail in \cite{bonnor1989radiating},\cite{chakraborty2024arrow}) 
\begin{equation} [p = Q]_{\Sigma},\label{matching}\end{equation}
which leads to the evolution equation for $a(t)$ (see details in \cite{chakraborty2024arrow} and \cite{bonnor1989radiating}):
\begin{equation}2a\,a_{,tt} +a_{,t}^2-2\alpha a\,a_{,t}=0,\qquad a_{,t}=-\frac{2\alpha}{\beta\,a^{1/2}}+ 2\alpha,\label{eqsS}\end{equation}
where $\alpha=[A_0'/B_0]_{\Sigma}$ follows from the condition $[p_0=0]_\Sigma$ at the initial static state. It is evident from \eqref{eqsS} (see detail in \cite{chakraborty2024arrow} and \cite{bonnor1989radiating}) that $a(t)$ decreases monotonically from an initial positive value. Hence, from equations (25)-(27) and (44)-(47) of \cite{chakraborty2024arrow}, the asymptotic time evolution of the variables $8\pi Q,\,8\pi \rho,\,8\pi p$ are respectively in decreasing order $ O( a^{-7/2}),\, O(a^{-3}),\, O(a^{-5/2})$, while $\Psi_2\sim O(a^{-2})$. Since $a(t)$ is rapidly decreasing, we have $(8\pi Q)^2\gg (8\pi \rho)^2\gg (8\pi p)^2\gg (\Psi_2)^2$, 
which implies $R_{ab}R^{ab}\gg C_{abcd}C^{abcd}$ and thus the ratio \eqref{ratioWR2} clearly shows a monotonically decreasing ${\cal P}\ll1$ with Ricci curvature becoming dominant over the Weyl-curvature.


\section{The gravitational arrow in a cosmological context}\label{sec:gravarrow}

In all previous work modeling the collapse of heat conducting spheres using the solutions \eqref{metric1}-\eqref{Tab}, the time evolution is completely determined by the matching conditions (including \eqref{matching}) with Vaidya that yield \eqref{eqsS}. However, given the strong geometric constraints posed by demanding zero shear, it is evident (and is acknowledged by the authors) that these simple toy models do not comply with physically motivated equations of state and appropriate heat transport equations linking the state variables $\rho,\,p$, and the dissipative flux $q_a$.  

To consider the shear-free solutions \eqref{metric1}-\eqref{Tab} as cosmological models not subjected to a matching with Vaidya, we need to provide a physical interpretation of the energy flux $q_a$ that is different from heat conduction, which is inappropriate in a cosmological context dominated by long range gravitational interaction. An alternative and potentially plausible physical interpretation is to regard $q_a$ in connection with a peculiar velocity field that results from the Lorentzian boost relating two distinct 4-velocities $\hat u^a$ and $u^a$.  
\begin{equation}\hat u^a =\gamma(u^a + v^a),\qquad \gamma = \frac{1}{\sqrt{1-v_av^a}},\quad v_a u^a =0. \label{boost}\end{equation}
where we assume $\hat u^a$ to be associated with the perfect fluid
\begin{equation} \hat T^{ab} = (\hat \rho+\hat p) \hat u^a \hat u^b + \hat p g^{ab},\label{hatPF}\end{equation} 
that is not comoving with the frame of $u^a$. Applying \eqref{boost} to \eqref{hatPF}, the energy-momentum tensor in the frame of $u^a$ is given by
\begin{eqnarray} 
T^{ab} &=& (\rho+p) u^a u^b + p g^{ab} + 2 q^{(a}u^{b)}+ \Pi^{ab},\label{nohat},\\
\rho &=& \hat \rho+\gamma^2 v_av^a (\hat \rho + \hat p),\qquad p= \hat p+\frac13\gamma^2 v_av^a (\hat \rho + \hat p),\label{hatrhop}\\
q^a &=& \gamma^2 (\hat \rho+\hat p) v^a,\qquad \Pi^{ab}=\gamma^2 (\hat \rho+\hat p) v^{\langle a}v^{b\rangle},\label{hatQPi}
\end{eqnarray}
where $v^{\langle a}v^{b\rangle}$ is the spacelike symmetric traceless tensor product. For non-relativistic peculiar velocities we have $v_av^a\ll 1,\,v^{\langle a}v^{b\rangle} \ll 1$ and thus $\gamma\approx 1+O(v^2)$, leading in linear order in $v^a/c$ to an energy-momentum tensor that is identical to \eqref{Tab}   
\begin{equation}T^{ab}=(\hat\rho +\hat p)u^au^b+p g^{ab} + 2q^{(a}u^{b)},\qquad q^a =(\hat \rho+\hat p) v^a,\label{nohatTab}\end{equation}
but now endowed with an interpretation that is consistent with a cosmological context. To apply \eqref{boost}-\eqref{nohatTab} to the models \eqref{metric1}-\eqref{Tab}, we can identify $\rho$ and $p$ in \eqref{Qrhop} with $\hat \rho$ and $\hat p$ in \eqref{hatPF}, using $\rho$ and $q_a$ to determine the peculiar velocity $v^a$. However, a full treatment is outside the scope of this paper and will be pursued elsewhere, as our purpose in presenting equations \eqref{boost}-\eqref{nohatTab} is to highlight the potentially plausible physical interpretation of these models. 

To show that the ratio \eqref{ratioWR2} need not be a monotonically decreasing function in the models \eqref{metric1}-\eqref{Tab}, we need to provide a guideline to determine their time evolution as inhomogeneous generalizations of spatially flat FLRW models, an approach justified by noting that the terms involving $a(t)$ and its derivatives in \eqref{Qrhop} are formally identical to the dynamical equations for the scale factor $a(t)$ of these models. However, the range of radial coordinate in \eqref{metric1} is now infinite, which can be problematic if the functions $\rho_0(\chi)$ and $p_0(\chi)$ in \eqref{Qrhop} diverge as $\chi\to \infty$. 

Following this approach and to prevent possibly divergent functions $\rho_0(\chi)$ and $p_0(\chi)$ as $\chi\to \infty$, it is convenient to generalize \eqref{metric1} to a metric analogous to FLRW models with positive spatial curvature  
\begin{eqnarray}ds^2=-\frac{N_0^2}{L_0^2} dt^2 + \frac{a^2 [d\chi^2+K^{-1}\sin^2 (\sqrt{K}\chi)(d\theta^2+\sin^2\theta d\phi^2)]}{L_0^2},\label{metric2}\end{eqnarray}
where $L_0=1/B_0>0$ and $N_0=A_0/B_0>0$ and $K>0$ is a constant in units of inverse length squared related to the spatial curvature: ${}^3R_0= 6K/a_0^2 $, with $a(t_0)=a_0=1$ (in what follows we omit these constants). 

Solutions of the field equations for the metric functions $N_0,\,L_0$ follow from the condition of pressure isotropy $G^\chi_\chi=G^\theta_\theta$ evaluated with the coordinate rescaling $\sin^2(\chi)= y(2-y)$ (or equivalently $y=1-\cos(H_0 \chi)$), leading to the following coupled second order linear differential equations
\begin{equation} \frac{d^2N_0}{dy^2}- J(y) N_0=0,\qquad 2\frac{d^2L_0}{dy^2}- J(y) L_0=0,\label{Jdef} \end{equation}
where $J(y)$ is an arbitrary function related to the conformal invariant $\Psi_2$ (it coincides with Equation (33) of \cite{chakraborty2024arrow})
\begin{equation} J=\frac{6H_0^2 \Psi_2 a^2}{y(2-y) L_0^2}\qquad \Rightarrow\quad \Psi_2 =\frac{J\,H_0^2\sin^2(\chi) L_0^2}{6a^2 }.\label{Psi2} \end{equation}
To find an exact analytic solution of \eqref{Jdef} we choose $J=\Delta^2$ in \eqref{Jdef}, with $\Delta>0$ an arbitrary constant, leading to 
\begin{equation}\fl N_0 = \cosh (\Delta y)+\nu\sinh (\Delta y),\qquad L_0=\cosh\left(\frac{\Delta}{\sqrt{2}}y\right)+\lambda \sinh\left(\frac{\Delta}{\sqrt{2}}y\right),\label{NLfun}\end{equation} 
where $y(\chi)=1-\cos (\chi)$ and $\nu,\,\lambda$ are integration constants that are fixed by demanding that $L_0=N_0=1$ hold at $\chi=\pi$:
\begin{equation}\lambda=\frac{1-\cosh (\sqrt{2}\Delta)}{\sinh (\sqrt{2}\Delta)},\qquad \nu = \frac{1-\cosh (2\Delta)}{\sinh (2\Delta)},\label{nulam}\end{equation}
 The dimensionless field equations equivalent to \eqref{Qrhop} are in terms of the metric functions 
\begin{eqnarray}
\fl \frac{8\pi\rho}{3H_0^2}=\frac{1}{H_0^2}\left[\frac{\dot a^2}{a^2}+\frac{k L_0^2}{a^2}\right]+\frac{1}{a^2}\left[\left(\frac13\Delta^2 L_0^2-L_0'^2\right)\sin^2 r +2\cos r L_0 \frac{dL_0}{dy}\right],\label{eqrho}\\
\fl \frac{8\pi p}{H_0^2}=-\frac{1}{H_0^2}\left[\frac{\dot a^2}{a^2}+\frac{2\ddot a}{a}+\frac{k L_0^2}{a^2}\right]\nonumber\\
\fl +\frac{1}{a^2}\left[\left(3\frac{dL_0}{dy}-2L_0\frac{dN_0}{dy}\right)\frac{dN_0}{dy}\sin^2 r+2\left(L_0\frac{dN_0}{dy}-2\frac{dL_0}{dy}\right)\cos r\right],\label{eqp}\\
\fl \frac{8\pi Q}{H_0}=\frac{2\dot a}{a}\left[\frac{L_0'}{L_0}-\frac{2N_0'}{N_0}\right],\qquad 64\pi^2q_aq^a=\frac{4\dot a^2}{a^4}\left[\frac{L_0'}{L_0}-\frac{2N_0'}{N_0}\right]^2,\label{eqQ}
\end{eqnarray}
where $y=1-\cos \chi$, the primes denote $\partial/\partial \chi =K^{-1}\sin(\sqrt{K}\chi) \partial/\partial y$ and $\dot a,\,\,\ddot a$ are proper time derivatives: $\dot a=da/d\tau=(L_0/N_0) a_{,t}$ and $\ddot a=d^2a/d\tau^2=(L_0/N_0)^2 a_{,tt}$. 

Since $0\leq \chi \leq \pi$ is bounded and $\Delta$ is arbitrary, we can assume $\Delta\,y\ll 1$ (as well as \eqref{nulam}) to expand the metric functions $N_0$ and $L_0$ and state variables \eqref{eqrho}-\eqref{eqQ} up to order $\Delta^2$, transforming \eqref{metric2} into the following near-FLRW metric 
\begin{eqnarray}\fl ds^2=-\left(1-\frac12 \sin^2 \chi \Delta^2\right) dt^2+\left(1+\frac12 \sin^2 \chi \Delta^2\right)\,a^2(t)[d\chi^2+\sin^2\chi(d\theta^2+\sin^2\theta d\phi^2)].\nonumber\\\label{metric3}\end{eqnarray}
Since up to order $\Delta^2$ we have $\dot a\approx a_{,t}$ and $\ddot a\approx a_{,tt}$, the terms with $a$ and its derivatives in \eqref{eqrho}-\eqref{eqQ} become standard FLRW equations for the scale factor $a$. In particular, we choose these equations to correspond to a cold dark matter source (dust) and a nonzero $\Lambda$ term, leading to 
\begin{eqnarray} 
\fl \frac{1}{H_0^2}\frac{a_{,t}^2}{a^2}=\frac{\Omega_0^m}{a^3}-\frac{\Omega_0^k}{S^2}+\Omega_0^\Lambda,\qquad -\frac{1}{H_0^2}\left[\frac{a_{,t}^2}{a^2} + \frac{2a_{,tt}}{a}+\frac{\Omega_0^k}{a^2}\right]=-3\Omega_0^\Lambda,\label{LCDM}
\end{eqnarray}
where $H_0$ is the Hubble radius, while $\Omega_0^m,\,\Omega_0^\Lambda$ and $kH_0^{-2}=\Omega_0^k=\Omega_0^m+\Omega_0^\Lambda-1>0$ are the standard observational parameters. Expanding the state variables \eqref{eqrho}-\eqref{eqQ} up to order $\Delta^2$ leads to 
\begin{eqnarray}
\fl \frac{8\pi\rho}{3H_0^2} = \frac{\Omega_0^m}{a^3}+\Omega_0^\Lambda +\frac{\left[\left(3\Omega_0^mS^3-2(3\Omega_0^m+3\Omega_0^\Lambda-1)S-3\Omega_0^m \right)\sin^2\chi-6S\right]\,\Delta^2}{6 a^2}, \label{rhoappr}\\
\fl \frac{8\pi p}{3H_0^2}=-\Omega_0^\Lambda=\frac{\left[(3a^2-2)\Omega_0^\Lambda+2(1-\Omega_0^m)\right]\,\sin^2\chi}{6a^2},\label{pappr}\\
\fl \frac{8\pi\,Q}{3H_0^2}=\frac{\left[\Omega_0^m-(\Omega_0^m+\Omega_0^\Lambda-1)a+\Omega_0^\Lambda a^3\right]^{1/2}\sin \chi \cos\chi \Delta^2}{3 a^{3/2}},\label{Qappr}\\
\fl \frac{64 \pi^2}{9 H_0^4}q_aq^a =\frac{1}{a^2} \left[\frac{8\pi\,Q}{3H_0^2}\right]^2\sim \frac{\Delta^4}{a^5},\qquad \Psi_2 =-\frac{H_0^2 \sin^2\chi \Delta^2}{6a^2 }\label{qqappr}
\end{eqnarray}
where only the variables common to FLRW ($\rho$ and $p$) have zeroth-order terms that coincide with their FLRW limit at both centers of symmetry $\chi=0,\,\pi$. We can now evaluate (up to order $\Delta^2$) the evolution of the Weyl to Ricci ratio \eqref{ratioWR2}. As we show in what follows, this ratio exhibits a very different behavior from that in the radiating spheres matched to Vaidya.

After substitution of \eqref{rhoappr}-\eqref{qqappr} in ${\cal P}$ in \eqref{ratioWR2} and expanding on $\Delta$ we obtain the following dimensionless quotient up to order $\Delta^4$ 
\begin{equation}{\cal P}\approx H_0^4\,\frac{\Delta^4\,\sin^4 \chi\,a^2}{4[\Omega_0^\Lambda]^2\,a^6+2\Omega_0^m\Omega_0^\Lambda\,a^3+[\Omega_0^m]^2} +O(\Delta^6),\label{PPappr} \end{equation}
which shows that ${\cal P}$ is an inhomogeneous quantity that vanishes at the two symmetry centers $\chi=0,\,\pi$ where FLRW conditions are expected. 

Although the ratio in \eqref{PPappr} was obtained from a series expansion, it does provide an accurate qualitative description of the evolution of ${\cal P}$ for the class of models under consideration, now described as positively curved near FLRW models. This ratio initially increases as $a^2$ from $a=0$ (initial big bang) for all parameter choices, but their subsequent evolution depends on $\Omega_0^m$ and $\Omega_0^\Lambda$. If $\Omega_0^\Lambda>0$ then whether the model perpetually expands or re-collapses (if $\Omega_0^m \sqrt{\Omega_0^\Lambda}< 2[(\Omega_0^m+\Omega_0^\Lambda)/3]^{3/2}$), there must be a cosmic time at which the $\Lambda$ term becomes dominant, reversing the growth of ${\cal P}$, which decays asymptotically to zero. For purely expanding models with $\Omega_0^\Lambda=0$ the ratio ${\cal P}$ grows as $a^2$ for all cosmic times. However, for re-collapsing models with $\Omega_0^\Lambda=0$ this ratio only grows in the expanding phase, decreasing to zero in the collapsing phase. 

\begin{figure}[h]
  \centering
  \includegraphics[width=0.8\linewidth]{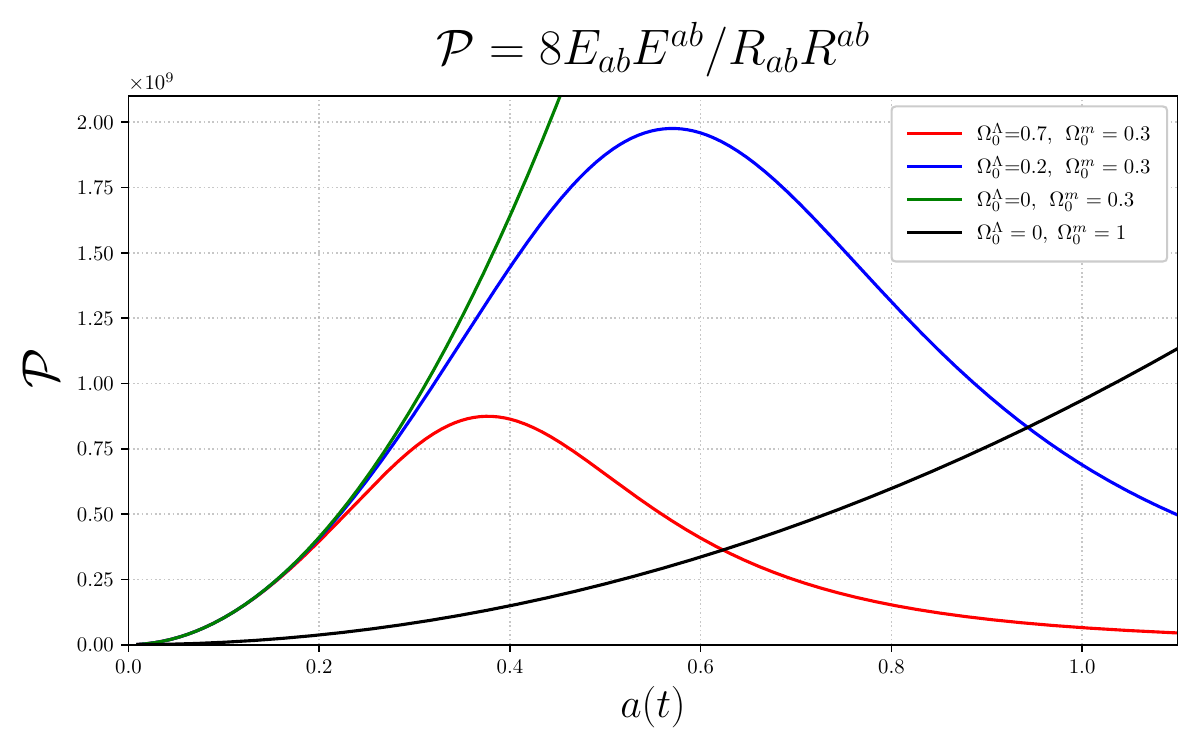}
  \caption{The Weyl to Ricci ratio $\mathcal{P}$ as a function of $a(t)$ with parameters 
  $\Delta = 0.1$, $\chi=\pi/2$, with different choices of $\Omega_0^m$ and $\Omega_0^\Lambda$, see figure labels for specific values. See main text for details.}
  \label{fig:PP_varOmL}
\end{figure}

\begin{figure}[h]
  \centering
  \includegraphics[width=0.8\linewidth]{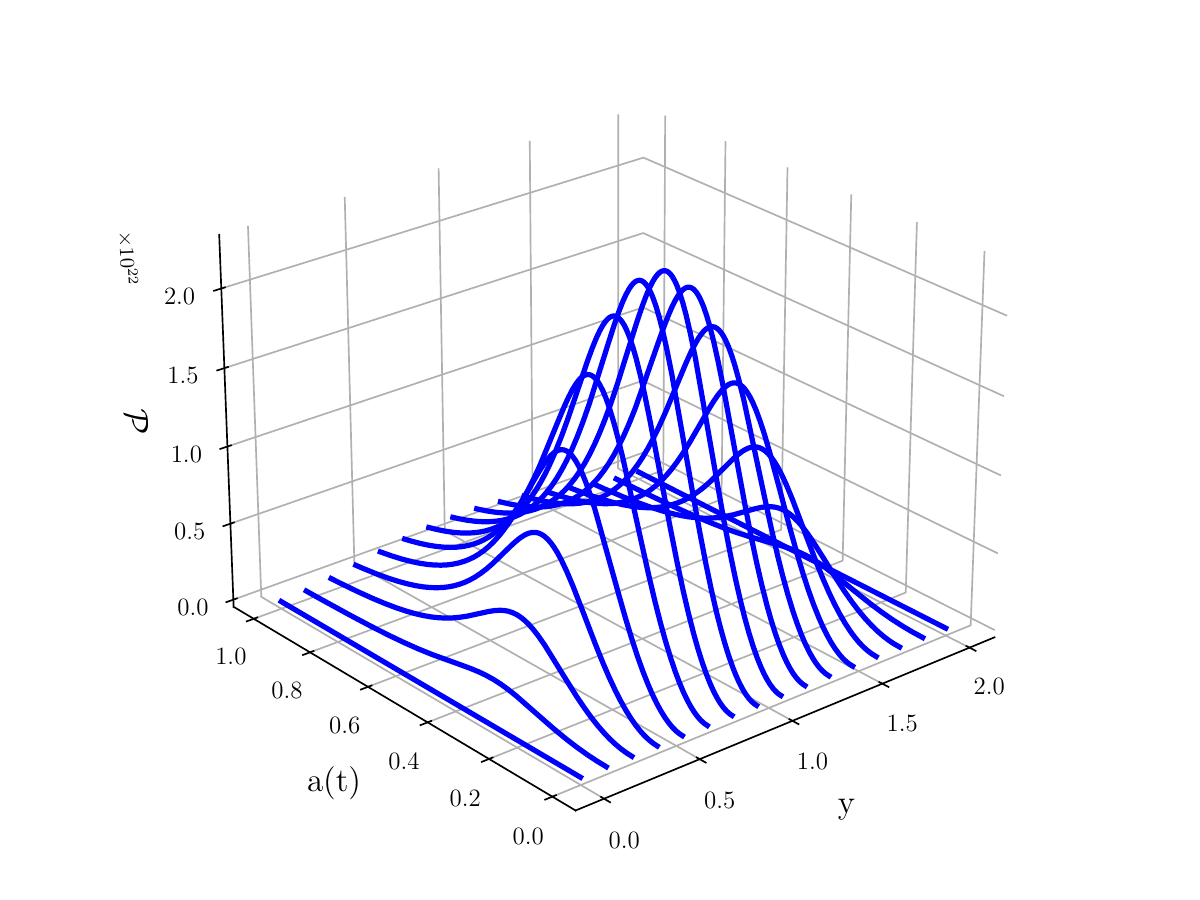}
  \caption{The Weyl to Ricci ratio $\mathcal{P}$ as a function of $a(t)$ and $y(\chi)$ for the model with  
  $\Delta = 0.1$, $\Omega_0^m = 1.05$, and $\Omega_0^\Lambda = 0.7$. See main text for details.}
  \label{fig:PP_3D}
\end{figure}

\begin{figure}[h]
  \centering
  \includegraphics[width=0.8\linewidth]{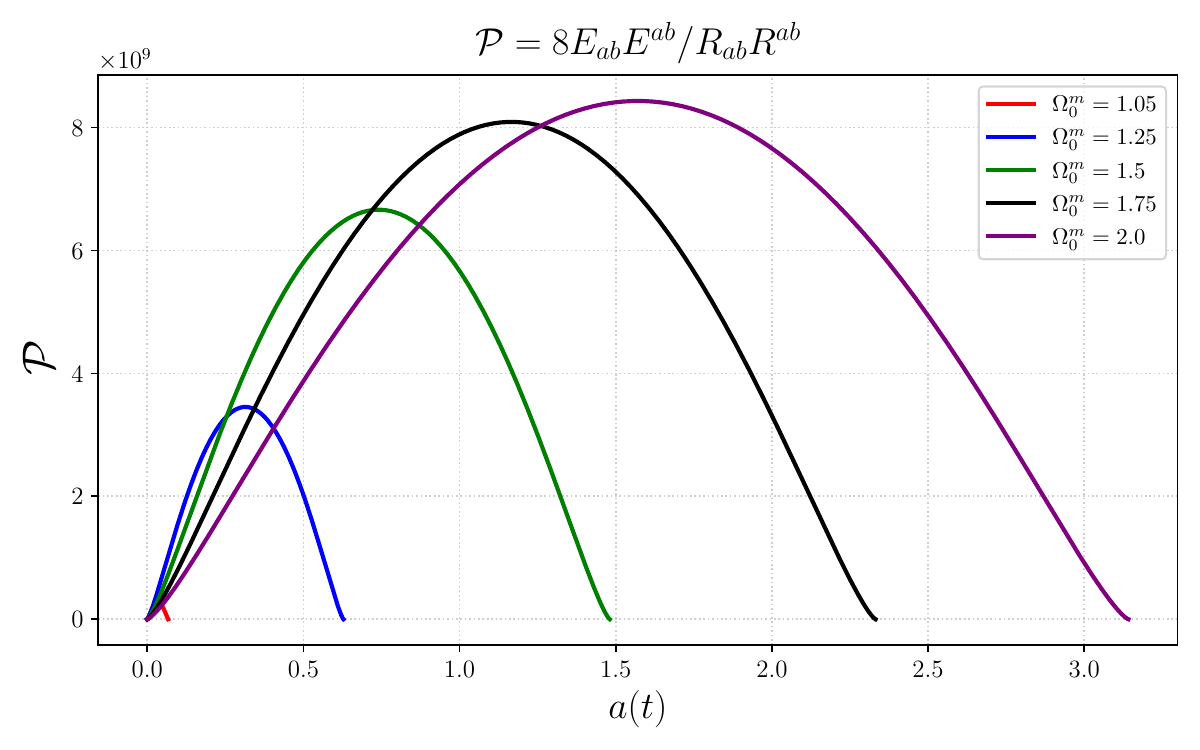}
  \caption{The Weyl to Ricci ratio $\mathcal{P}$ as a function of $a(t)$ for the model with 
  $\Delta = 0.1$, $\chi=\pi/2$, $\Omega_0^\Lambda = 0$, and various choices of $\Omega_0^m$. See main text for details.}
  \label{fig:PP_noOmL}
\end{figure}

Figures \ref{fig:PP_varOmL} and \ref{fig:PP_3D} graphically illustrate this qualitative description (we take $\Delta=0.1$ in all plots). The parameter combination $\Omega_0^m=0.31,\,\, \Omega_0^\Lambda=0.7$ corresponds to a model with very small positive spatial curvature ($\Omega_0^K=0.01$) that approximates a $\Lambda$CDM model consistent with observations. We plot in Figure \ref{fig:PP_3D} the ratio ${\cal P}$ for this model as a function of $r$ and $a$, clearly showing the described pattern. To illustrate the effect of an increasing $\Lambda$ term, we plot in Figure \ref{fig:PP_varOmL} the evolution of ${\cal P}$ along the worldline $\chi=\pi/2$ for models with $\Omega_0^m=0.31$ and a sequence of values $\Omega_0^\Lambda=0.5,\,0.7,\,1.0,\,1.5,\,2.5$, showing how the maximum of ${\cal P}$ decreases for increasing $\Omega_0^\Lambda$, in consistency with its role in suppression of structure formation. As a comparison, we also plot the case $\Omega_0^\Lambda=0$ that increases monotonically as $a^2$. Models with $\Omega_0^\Lambda=0$ and increasing values $\Omega_0^m=1.05,\,1.25,\,1.5,\,1.75,\,2.0$ are plotted in Figure \ref{fig:PP_noOmL}, showing a maximum of ${\cal P}$ at larger values of $a$ with increasing $\Omega_0^m$. 

It is interesting to compare the evolution of the Weyl to Ricci ratio ${\cal P}$ for the models under consideration with the corresponding evolution of the gravitational entropy proposal of Clifton, Ellis and Tavakol (CET) \cite{clifton2013gravitational}, based on an effective energy-momentum tensor constructed from the Bel-Robinson tensor. For Petrov type D spacetimes, the CET entropy production $\dot S_{_\textrm{\tiny{gr}}}$ is given in terms of the ``gravitational'' energy ${\cal E}_{_\textrm{\tiny{gr}}}$ and temperature $T_{_\textrm{\tiny{gr}}}$ by
\begin{eqnarray}
\dot S_{_\textrm{\tiny{gr}}}=\frac{\dot {\cal E}_{_\textrm{\tiny{gr}}}}{T_{_\textrm{\tiny{gr}}}},\qquad  {\cal E}_{_\textrm{\tiny{gr}}}=|\Psi_2|V,\qquad T_{_\textrm{\tiny{gr}}} =\left| \nabla_{_\textrm{\tiny{A}}} u_{_\textrm{\tiny{B}}} k^{_\textrm{\tiny{A}}}\,l^{_\textrm{\tiny{A}}}\right|
\end{eqnarray}
where the dot is proper time derivative, $V$ is a local 3-dimensional volume (such that $\Theta=u^a\,_{;a}=\dot V/V$), while $k^a,\,l^a$ are null vectors associated with an orthonormal tetrad. Since $T_{_\textrm{\tiny{gr}}}\geq 0$ by constructions, the sign of $\dot S_{_\textrm{\tiny{gr}}}$ is determined only by the sign of $\dot {\cal E}_{_\textrm{\tiny{gr}}}$. For the models under consideration $V= a^3 \sin^2\chi/L^3$ and using \eqref{Psi2} we have 
\begin{equation}{\cal E}_{_\textrm{\tiny{gr}}} = \frac{H_0^2 \Delta^2 \sin ^4\chi}{6L_0}\,a\quad\Rightarrow\quad \dot {\cal E}_{_\textrm{\tiny{gr}}}=\frac{H_0^2 \Delta^2 \sin ^4\chi}{6L_0}\,\dot a\end{equation}
where $\dot a=(L_0/N_0)a_{,t}$. Consequently, there is CET gravitational entropy growth ($\dot S_{_\textrm{\tiny{gr}}}\geq 0$) in all expanding stages of the models, while this entropy decreases in all collapsing stages ($\dot a<0$). This means that CET gravitational entropy necessarily decreases in the models used by Bonnor \cite{BONNOR1987305}, and \cite{bonnor1989radiating,chakraborty2024arrow} to describe collapsing spheres in a Vaidya exterior, since $a(t)$ in \eqref{eqsS} is monotonically decreasing. The CET entropy also decreases in the Vaidya exterior, since (from \eqref{vaydia2}) $|\Psi_2|V= M(u)$ is a positive but decreasing function. 

\section{Closing remarks}\label{sec:closing}

To summarize: for the shear-free models under consideration the gravitational arrow based on the Weyl to Ricci ratio ${\cal P}$ in \eqref{ratioWR2} and the CET entropy $\dot S_{_\textrm{\tiny{gr}}}$ both decrease in collapsing configurations, but ${\cal P}$ decreases in expanding models with $\Omega_0^\Lambda>0$ for which $\dot S_{_\textrm{\tiny{gr}}}$ would increase. Evidently, these simple patterns follow from the extremely simplified way we have defined the time evolution of $a(t)$ using the FLRW equations (whose collapse is time-symmetric with the expansion) in models not subjected to a matching with Vaidya (whose matching conditions provide a rigorous procedure to define the function $a(t)$ in \eqref{eqsS}). In fact, in LTB dust models the collapsing stage is not time-symmetric with respect to the expanding one, hence $\dot S_{_\textrm{\tiny{gr}}}$ increases in this more realistic collapse scenario \cite{sussman2014gravitational, sussman2015gravitational}.

We have proved that the gravitational arrow of time based on the Weyl to Ricci ratio ${\cal P}$ is not monotonically decreasing in expanding cosmological models of the same class of shear-free solutions examined in \cite{chakraborty2024arrow, BONNOR1987305, bonnor1989radiating} as collapsing spheres in a Vaidya exterior. We have also provided arguments supporting a physically plausible interpretation of these solutions as cosmological models admitting a field of peculiar velocities, an appealing property that will be further studied elsewhere in forthcoming work.     

\section*{Acknowledgements}
SN acknowledges support form Postdoctoral grant from Secretar\'ia de Ciencia, Humanidades, Tecnolog\'ia e Innovaci\'on SECIHTI. JCH acknowledges support from the UNAM-PAPIIT grant IG102123 "Laboratorio de Modelos y Datos (LAMOD) para proyectos de Investigaci\'on Cient\'ifica: Censos Astrof\'isicos", as well as from SECIHTI (formerly CONAHCYT) grant CBF2023-2024-162 and DGAPA-PAPIIT-UNAM grant IN110325 "Estudios en cosmolog\'ia inflacionaria, agujeros negros primordiales y energ\'ia oscura."

\section*{References}
\bibliographystyle{iopart-num}
\bibliography{references}

\end{document}